# Animal social networks: an introduction for complex systems scientists

Josefine Bohr Brask*[1,2], Samuel Ellis[1], and Darren P. Croft[1]

[1]*Centre for Research in Animal Behaviour, College of Life and Environmental Sciences, University of Exeter, Exeter, UK*

[2]*Department of Applied Mathematics and Computer Science, Technical University of Denmark, Lyngby, Denmark*

*Corresponding author. Email: bohrbrask@gmail.com

Many animals live in societies where individuals frequently interact socially with each other. The social structures of these systems can be studied in depth by means of network analysis. A large number of studies on animal social networks in many species have in recent years been carried out in the biological research field of animal behaviour and have provided new insights into behaviour, ecology, and social evolution. This line of research is currently not so well connected to the field of complex systems as could be expected. The purpose of this paper is to provide an introduction to animal social networks for complex systems scientists and highlight areas of synergy. We believe that an increased integration of animal social networks with the interdisciplinary field of complex systems & networks would be beneficial for various reasons. Increased collaboration between researchers in this field and biologists studying animal social systems could be valuable in solving challenges that are of importance to animal social network research. Furthermore, animal social networks provide the opportunity to investigate hypotheses about complex systems across a range of natural real-world social systems. In this paper, we describe what animal social networks are and main research themes where they are studied; we give an overview of the methods commonly used to study animal social networks; we highlight challenges in the study of animal social networks where complex systems expertise may be particularly valuable; and we consider aspects of animal social networks that may be of particular interest to complex systems researchers. We hope that this will help to facilitate further interdisciplinary collaborations involving animal social networks, and further integration of these networks into the field of complex systems.

## 1 Introduction

Animals of many species live in groups where individuals spend time in close proximity to each other and frequently interact [1]. The patterns of social interactions and spatial proximity across individuals constitute the social structures of the populations. During the last two decades, scientists in the biological research field of animal behaviour (and particularly in the subdiscipline of behavioural ecology) have investigated the social structures of a wide range of species by means of network analysis (reviewed in [2]). While the network approach has been used in animal behaviour research for many years (in particular primatology [3]), advanced quantitative network analysis tools started to become generally adopted in the field of animal behaviour in the beginning of this millennium [3–8], and social network analysis has since become a well-integrated and widely applied part of animal behaviour research that continues to provide new insights into diverse questions about sociality, ecology and evolution [2, 9, 10]. A large amount of research on animal social networks has by now been conducted within the animal behaviour field, including empirical studies of animal social networks, network-based modelling of animal social systems, and development of analytical methods specifically designed for these networks [2, 6, 10, 11]. Animal social networks have also been studied by complex systems researchers (and researchers working in both the animal behaviour and complex systems fields), and studies of the networks have been published outside of biological journals (e.g. [12–16]). However, the research on animal social networks going on in the field of animal behaviour is overall less extensively connected to the field of complex systems than might be expected, despite strong connections between the animal behaviour field and the complex systems field [17-22].

The purpose of this paper is to provide an introduction to animal social networks for the wider complex systems & networks research community, and to suggest scientific challenges and aspects in relation to these networks that may be of particular cross-disciplinary interest and where synergy of the two fields could be particularly profitable. Our hope is that this can facilitate further integration of this area of research into the interdisciplinary field of complex systems & networks, and strengthen the connection between biologists studying animal social networks and researchers with expertise in complex systems. We believe that this would benefit both our understanding of animal social systems and the general research in complex systems for various reasons. Animal social network research is facing specific challenges where input from computational and theoretical scientists with knowledge about complex systems can be highly relevant for finding good solutions. Overcoming these challenges is a relevant scientific endeavour because animal social networks constitute a class of networks that play a central role in evolutionary and ecological processes [2, 9, 10], and they are therefore important to study in their own right. Furthermore, similarly to empirical networks from other domains, animal social networks may be used in the general study of complex systems, and they provide data from a wide range of natural social systems and offer varied possibilities for experimental manipulation of the systems. In the following we seek to provide information that may be of relevance both for complex systems researchers new to the topic of animal social networks, and those that are well acquainted with it and are seeking further information or inspiration. We have aimed to keep the paper relatively brief and there will naturally be some subjectivity in what is included, but we hope that it may be found useful.

The paper is structured as follows: We first briefly explain what animal social networks are (Section 2). We then provide overviews of some main themes for animal social network research (Section 3), and of methods commonly used in studies of animal social networks (Section 4). We thereafter describe some current challenges for research in these networks where complex systems expertise may be particularly valuable (Section 5), and aspects of animal social networks that may be of particular interest for complex systems researchers (Section 6). We finish with a note on the availability of animal social network data (Section 7), and a brief conclusion (Section 8).

## 2 What are animal social networks?

Here we provide a brief explanation of what animal social networks are - what kind of data they represent and what types of patterns are typically observed in them. For further information about common themes and methods for animal social network studies, we refer the reader to Section 3 and 4 and references therein.

Animal social networks quantify the social structure within animal populations (see Fig. 1 for examples of animal social network graphs). Each node in the network corresponds to a specific individual, and the (typically weighted) network edges correspond to the social relationships between the individuals, which are quantified as rates of *social interaction* or *social association* between each dyad [2, 6, 10, 27]. Social interactions commonly used for quantifying animal social structure include grooming and fighting, whereas social associations are based on spatial proximity of individuals. The networks may thus quantify different dimensions of the social system, depending on what type of social interaction (affiliative, aggressive. . . ) or social association they are based on. The network data (the adjacency matrix) will often be accompanied by attribute data, which usually contain information on the individuals (their sex, age, body size, etc.) or the dyads (e.g. their genetic relatedness).

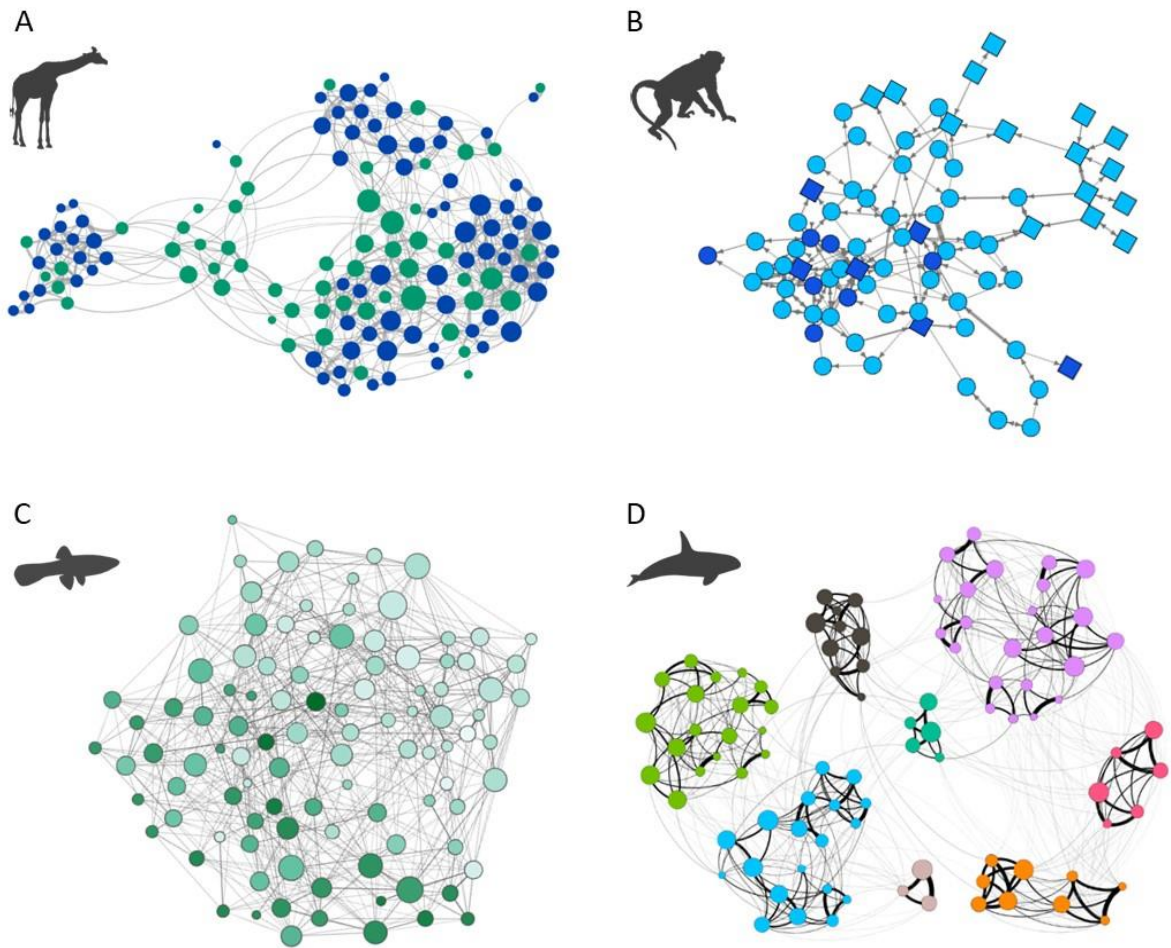

**Figure 1.** Examples of animal social network graphs. In all graphs, each node signifies an individual. All graphs are weighted, with thicker edges signifying a stronger social relationship (corresponding to a higher rate of social interaction or association). A) A social network of a population of giraffe, based on social associations (individuals observed in the same group). Females are shown in blue and males in green, larger nodes have higher degree, and very weak edges are not shown for clarity. B) A social network of a group of rhesus macaques, based on grooming interactions. Females are shown as circles and males as squares, darker node colour indicates high-ranking individuals, and edges point from the groomer to the individual being groomed. C) A social network of a population of Trinidadian guppies, based on social associations (individuals observed shoaling together). Larger nodes have higher degree and darker nodes signify individuals with a larger body size. D) A social network of a killer whale population, based on social associations (individuals observed in the same group). Node colour indicates network communities and larger nodes have higher within-community closeness. The macaque graph is reproduced from [23] and the killer whale graph is modified from [24] with permission from the authors (http://creativecommons.org/licenses/by/4.0/; killer whale silhouette by Chris Huh, https://creativecommons.org/licenses/by-sa/3.0/). The giraffe and guppy graphs are previously unpublished; for data collection methods see respectively [25] and [26].

Animal social networks are studied both in wild and captive populations. While field data enables the study of animal social structures under natural conditions, laboratory-based studies allow for experiments where causality can be tested under controlled conditions. In both cases, the quantified networks are most often relatively small ($N < 200$ [28]). They are typically studied by means of statistical analyses and other quantitative techniques.

By now, animal social networks have been quantified and analysed in a wide range of species, including mammals, birds, fish, reptiles and insects (reviewed in [2]). It is clear from this large body of research that social networks of many animal species are non-random, in the sense that their structures differ from what would be expected under random interaction or association. Typical patterns
3

observed in the networks include substantial variation in edge weights, pronounced modularity, and assortment by physical and behavioural individual characteristics (see Fig. 1 for examples). Such non-random structure is found across the taxa investigated (not only in species such as primates that have traditionally been considered more 'complex'), with high diversity of observed patterns [2]. Understanding the processes (evolutionary and proximate) underlying these structural patterns, and the implications of these network structures for social evolution, behaviour, and dynamics on the networks, is a central endeavour in the study of animal social networks.

## 3 Why are animal social networks studied?

Analyses of animal social networks are used in investigations of a wide range of questions about social evolution, behaviour and dynamical processes [2, 9, 10]. As we cannot cover all of these questions here, we instead consider some main research themes in animal social network research so far. It may be noted that the research themes overlap considerably with common themes in general complex systems & networks science, thus providing a natural base for further integration of animal social network research into this field.

**Social centrality, evolution and fitness.** A major reason why animal social networks are of scientific interest is that the social environment can act as an important driver of the evolution of traits (including both physical and behavioural characteristics of individuals [9, 29]). This means that in order to understand evolution, the social environment must be taken into account. Network analysis provides tools to quantify social structure in detail and across different scales, and therefore facilitates comprehensive studies of the role the social environment plays in evolution, across species (with a perspective that is complementary to that of quantitative genetics approaches [29]). One way to empirically investigate the evolutionary importance of the social environment is to statistically test for relationships between the social network positions of individuals and their Darwinian fitness (i.e. the extent to which they contribute to the future gene pool, which is commonly estimated by measures of longevity, reproduction rate, and offspring survival). In recent years, such studies have been carried out in a range of species, and evidence for correlations between fitness and various measures of network centrality has been found widely ([24, 30-36], for overviews see [37, 38]). The study of animal social networks is thus providing empirical evidence that social network position is linked to survival and reproduction across species. Fitness consequences of social position are furthermore increasingly being linked to broader themes such as the evolution of senescence and the evolution of cognition, and feedback between network position, traits, and fitness-related conditions (e.g. infection status) are being explored [29, 39-41].

**Spread of disease and information in networks.** Animal social networks are providing new insights into spreading processes in animal systems, including the propagation of disease and information. Studies in a range of species have investigated empirically to which degree various types of information spread via social links in both wild and captive populations, either via studying natural information or via experimental seeding of novel information. Types of information studied include the location of food [42-45], and innovations such as tool use [46, 47] and other new foraging techniques [48-52]. Regarding disease spread, empirical studies have uncovered relationships between individual network position and infection status or parasite load in multiple species [53-57], and simulation studies involving real-world animal social network data have given insights into the effect of social structure on disease transmission and the vulnerability of populations to epidemics [13, 58-61]. Animal social network data in combination with simulations have also been used to investigate more general aspects of spreading processes (e.g. [14]).

**Stability, flexibility and robustness of social systems.** Another area where animal social network research is providing new knowledge concerns the stability of social structures across species, and how flexible and robust they are under changing conditions and perturbations. This is studied by investigations of to which degree animal social network structures are stable across years [25, 62-66], how the network structures correlate with environmental factors such as food availability [67, 68] and general seasonal changes [25, 69-72], and how they react to node loss (see *Network robustness*, Section 5).

**Cooperation in structured populations.** Animal social networks are also used in studies concerning the evolution of cooperation. Cooperative behaviour - acting to benefit others at a cost to yourself - is studied widely across scientific fields, and theoretical and modelling-based work, as well as experiments with humans, have shown that social network structure can have important effects on the survival of cooperative behaviour [73-76]. The role that social structure plays in the evolution of cooperation in real-world systems in nature is, however, currently not well understood. Studies linking real-world animal social network structures and cooperative behaviour can provide insights into to which extent and under which conditions theoretically predicted structural effects are important for the maintenance of cooperation in social systems in nature. Research connecting cooperation and animal network structures conducted so far includes simulations of the spread of cooperation in animal social networks [77, 78], and empirical investigations of animal social network structure in relation to, respectively, cooperative courtship displays [79-82], food sharing [83], and social assortment by cooperativeness [26].

**Wildlife conservation and animal welfare.** The fact that social network structure has important implications for health, survival and behaviour across species means that animal social network studies have an important role to play in the conservation of wildlife [40, 84] and in improving the welfare of farm and zoo animals [85, 86], thus providing important drivers for applied animal social network studies. Such studies are for example concerned with estimation of the efficiency of disease control strategies in endangered wildlife [87-89], assessment of social behaviour in connection with relocation or reintroduction of animals into the wild [90, 91], and informing the management of captive populations [92].

## 4  How are animal social networks studied?

The study of animal social networks involves a combination of general network analysis approaches, and procedures developed specifically for these networks. In this section we give a brief overview of the methodology currently commonly used for data collection, network construction and network analysis in animal social network research, with reference to specialised literature for further details.

**Data collection.** The type of data collected to quantify animal social relationships (association or interaction data, see section 2) and the method of collection depends on the research question, what behaviour is possible to observe, and the social dynamics of the system. In species where group compositions change continuously on a relatively fast scale (e.g. within days or minutes depending on species), social association is often inferred from shared group membership (an approach known as *the gambit of the group* [93]), and the network data are collected by recording repeatedly over time which individuals are grouping together in space [27, 94, 95]. When groups are either largely stable across the observation period or group boundaries cannot easily be defined, then social association may be inferred from other criterions of spatial proximity [27, 96]. Interaction data may be collected when direct interactions between individuals occur frequently enough to estimate edge weights. The data may be collected either by the researchers directly observing the animals, or by automatic recording. In the former case, the researchers must be able to recognise each individual, which can be done by natural markings such as fur patterns and scars, or by equipping the animals with artificial tags. Methods for automatic data collection of animal social network data are becoming increasingly common due to the continuous optimisation of the involved technology (for detailed overviews see [97, 98]). Highly detailed data can be obtained via proximity loggers attached to each animal (Fig. 2), which - similarly to some methods used for human social networks - record when each pair of individuals are close to each other (for example [87, 99-102]); this can give datasets of social associations with a sub-second time resolution. The loggers may also contain other sensors, such as accelerometers, which can provide additional information on the behaviour of the animals. Another possibility is to use RFID tags to record when each animal is present at a specific location (for example [42]). Furthermore, high-resolution social association data can in some circumstances be obtained by simultaneous automatic tracking of multiple individuals from videos with methods based on machine learning (for example [103, 104]), either without tagging the animals or with computer-readable tags such as barcodes. The increase in the development and use of automatic data collection methods means that the future is likely to see high-resolution datasets of animal social networks across many species.

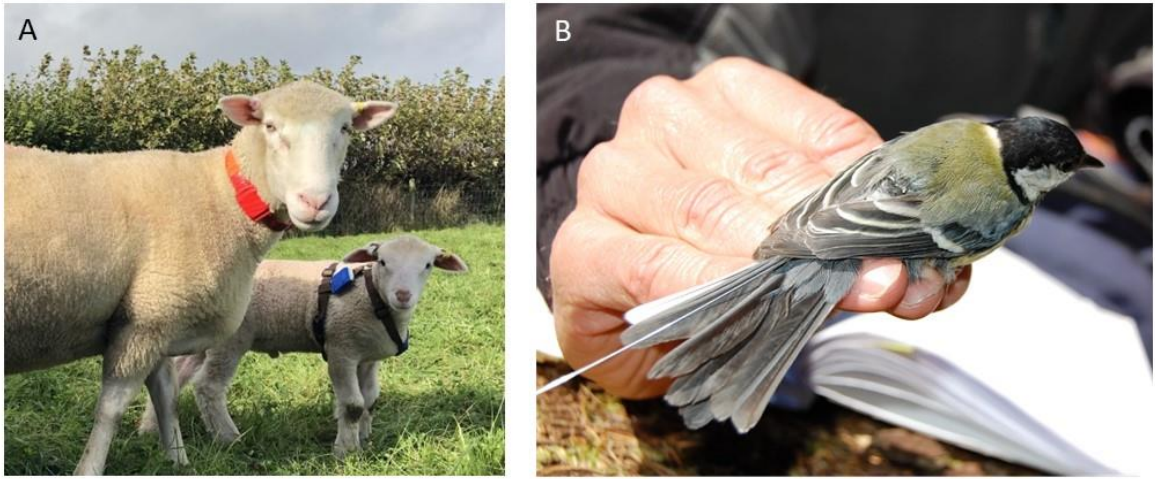

**Figure 2.** Examples of animals wearing electronic devices for collection of social network data via proximity sensing. A) Ewe and lamb wearing a collar and harness with devices attached (photo by Emily Price). B) Great tit wearing a miniature device with antenna on its back (photo by Lysanne Snijders).

**Network construction.** Edge weights in animal social networks are commonly estimated based on the observed rates of interaction or association for each pair of individuals. While the raw numbers of interactions of associations may sometimes be used directly as edge weights, most often edge weight estimations take into account a number of factors that could affect the amount of interaction or association but are not of interest for the research question [6, 105, 106]. These factors include different observability of individuals (e.g. from individuals being out of sight or from demographic changes during the data collection period), different observability of groups versus single individuals, and correlation in datapoints due to temporal closeness. For association data, standardised methods for the edge weight estimations are used in the form of *association indices* [6, 105, 106]. Before the calculation of edge weights, the data are often restricted by applying a threshold for the minimum number of times an individual should be observed in order to be included in the network, to decrease the amount of uncertainty on the edge weight estimates [6]. With the current increase in the use of automatic data collection methods in animal social network studies, some network construction issues become less relevant for these networks (e.g. high uncertainty on null associations [98]), while the new data formats require other considerations and development of suitable data extraction techniques (e.g. inferring spatiotemporal co-occurrences of individuals from data streams [107, 108]).

**Network analysis.** Studies of animal social networks often involve statistical analysis of the network structure, to test for structural patterns arising from biological processes [6, 27, 109]. The tests typically concern global structural patterns, correlations between network positions and individual attributes (sex, age, etc.), or correlations between edge weights and other dyadic data (e.g. relatedness, space use overlap, and social network data for the same set of individuals measured under other ecological or experimental conditions). Global network structure and network positions are quantified by standard network metrics (assortativity coefficients, centrality metrics etc.) and other similar measures. Statistical significance is typically tested by means of null models based on data permutation procedures (due to the non-independency in the data), which often
6

involves accounting for other factors that could lead to non-random patterns but are not of interest for the hypothesis, including for example differential observability of individuals, group size distribution, and demographic changes during the data collection. Identifying the appropriate null model for the research question is a key aspect of animal social network analysis [109-111], and the development of specialised data permutation procedures, and other control procedures, is an area of major interest and importance in animal social networks research. In addition to statistical testing for structural patterns, a range of other types of analyses have also been employed in animal social network research, such as robustness analysis (see Sections 3 and 5), community detection, network-based diffusion analysis (for information flow [112]), and methods for investigating disease spread [40, 113, 114]. In contrast to other domains of network science, degree distributions and their implications has not been a focus topic in relation to animal social networks, as their small size limits robust fitting of degrees to theoretical distributions [115, 116]. The methodology of animal social network analysis is continuously evolving and expanding, and going forward animal social network research is starting to explore and use additional methodological approaches, including relational event models [117], exponential random graph models [118], stochastic actor-oriented models [119], time-ordered networks [120], and multilayer networks [121]. Together this points towards increasingly diverse and multidimensional analyses of animal social networks.

## 5   Some challenges in animal social network research where complex systems expertise may be of particular advantage

There are many ways in which complex systems expertise can be useful for understanding animal social networks. In the following, we describe some challenges for animal social network studies where input from scientists with expertise in other types of empirical networks or in theoretical aspects of complex systems may be particularly valuable for finding good solutions.

**Network similarity.** An important challenge for animal social network research is how to measure the similarity between real-world networks from different sets of individuals in a meaningful way [11, 116]. Comparing the social structures of different species, or populations of the same species living in different environments or containing different compositions of individuals (e.g. with regard to sex or age), could potentially bring new key insights into the evolution of social systems and how they are shaped by internal and external factors. In animal social network research, network similarity is often investigated by quadratic assignment procedure matrix correlation methods [122], but these can only be used for networks that contain the same set of individuals (e.g. the same group under different environmental conditions). The comparison of animal social networks from social groups with different sets of individuals is complicated by the fact that the groups will often by necessity have different sizes. Furthermore, differences in data collection (Section 4) can lead to spurious differences in network structure [123]. Quantitative comparison of animal social networks from different individuals has therefore rarely been done and there is currently no general use of specific network similarity measures for these networks (although specific approaches have been suggested and used, e.g. motif analysis [11, 124] and exponential random graph models [118]). Given the fact that graph similarity is a fundamental topic of interest in complex systems & networks science, with a wide diversity of distance measures developed for various purposes [125], there should be much scope for interdisciplinary development of new network comparison methods that are specifically designed for animal social network data and capture differences of interest for these systems.

**Network complexity.** Another challenge of high relevance for research in animal social networks is the development of relevant measures of network complexity for these systems. Animal social complexity, and its variation between and within species, has long held interest from animal behaviour researchers, both because it provides a framework for understanding the evolution of social systems, and because of its potential links to the evolution of cognitive abilities and communication systems [17, 126, 127]. However, defining useful, detailed measures of social complexity is not straightforward, particularly if they are to be used across systems (different populations and species). The quantification of animal social systems as networks entails a potential for new, high-resolution measures of animal social complexity. This has not yet been much explored (for exceptions, see [128] for a recent suggestion for a network-based animal social complexity measure and [6] for a discussion of various potential measures; see also [17] for a general discussion of animal social complexity measures in relation to complex systems concepts). Collaboration between theoretical researchers with expertise in complexity quantification and empirical animal behaviour researchers seems very relevant for advancing the area of network-based complexity measures for animal social systems.

**Network robustness.** A topic which has got somewhat more attention and may also particularly benefit from increased interdisciplinary collaboration is the robustness of animal social networks, i.e. their ability to withstand the loss of network nodes (individuals). Knowledge about this is important for the conservation of animal populations (e.g. in the face of poaching or habitat destruction, which can lead to network fragmentation and reduction in network size), as well as for understanding social evolution and the role natural demographic changes play for social dynamics [129]. Robustness of animal social networks has been investigated by simulated [12, 130-132] and actual (experimental and natural) removal of individuals from the populations [12, 133-137]. Thus, we now have knowledge about the effects of node loss in multiple species, including real-world system reactions to such loss. Better integration of fundamental mathematical approaches for investigating network robustness (such as percolation theory and related areas) in combination with knowledge about the systems (including about system functionalities, and system anticipation and reaction to node loss [129]) could potentially lead to greater synthesis and general predictions about the robustness of social systems across species.

**Multilevel network structure.** A recent focus of attention in the study of animal behaviour is the extent to which animal societies can be considered 'multi-level'- that is, consisting of nested social levels (for an overview see [138]), and there is currently discussion in the field concerning how best to define and quantify multi-level structure in animal social systems [138-140]. Given the fact that hierarchical and nested structure has long been of focus in the field of complex systems & networks [141], collaboration concerning such structural patterns in animal social systems has the potential to be very valuable.

**Extraction of information from large datasets.** Finally, the automated data collection methods that are now in use (described in Section 4) means that animal social network datasets are increasingly large and multidimensional, and the extraction of information from the raw data is less direct. Optimisation of the treatment of these data is likely to benefit from interaction with areas of complex systems science where large and complex datasets are routinely dealt with.

## 6 Some aspects of animal social networks that could be of particular interest for complex systems researchers

Animal social networks may be - and are - used by complex systems researchers in a variety of ways. Available networks can be used as examples of real-world networks alongside networks from other systems (for example in studies developing new network methodology; e.g. [142]). Studies of animal social networks may also be designed specifically for investigating questions about complex systems [143]. These networks have features that make them unsuitable for some investigations (e.g. their relatively small size), and whether animal social network data is more or less useful than those from other real-world systems (including human social systems) will depend on the particular study and research question. We here highlight some aspects of animal social networks that - while not necessarily unique to these systems - may be of particular interest for complex systems researchers.

**Diversity of natural social systems.** Animal social systems of different species and populations are shaped by different selection pressures and ecological circumstances and characterised by strikingly different social dynamics and resulting network structures [2, 9, 10], ranging from structures based on family groups that are stable over multiple generations (e.g. killer whales, Fig. 1), to those based on groups that split and merge on a sub-minute time scale (e.g. Trinidadian guppies, Fig. 1). The diverse forms of sociality are of great scientific interest and social systems are studied across taxonomic groups, including insects, fish, mammals, birds and reptiles [2, 28]. Animal systems therefore constitute a source of network data from diverse natural social systems, which may be useful e.g. for understanding the evolution and behaviour of complex systems in nature, and datasets from a range of species are now publicly available (see Section 7; while this provides a choice of network data from different social systems, we note that direct comparison of networks from different species and populations comes with some caveats, see *Network similarity*, Section 5).

**Experimental manipulation of social networks.** Animal systems provide the opportunity to experimentally manipulate real-world social networks in a variety of ways and settings. Experimental treatments may be applied both in the wild and in the laboratory, with examples including establishment of experimental groups with different compositions of individuals (e.g. [144]) experimental temporary removal of individuals [12, 136] and experimental seeding of novel behaviours [49-51]. While laboratory-based experiments make it possible to test under controlled conditions how various factors affect network structure and dynamics, experiments in the field can provide insights into causal effects in natural populations.

**Replicated real-world networks.** For species that are easily kept in the laboratory or similar settings, researchers can establish replicated experimental social groups, from which social network data can be collected. For smaller species in particular, multiple groups of equal size can be established, allowing for statistical testing of differences in global network structure between experimental treatments (e.g. [144]).

**Social networks across lifetimes and generations.** Long-term behavioural research studies of wild animal populations have resulted in datasets of social interaction and association patterns that span lifetimes and generations, for a range of species. This allows for studies of multi-generational social networks in various real-world systems, including studies on long-term stability of network structures [25, 62-66], links between social environment and longevity [24, 32, 33, 35, 36], and network effects of demographic (node) turn-over [129]. Sociality across lifetimes and multiple generations may also be studied in the laboratory within relatively short time-frames in short-lived species.

**Modelling of social networks.** While generative network models have not had as central a role in animal social network research as in network science in general, such models have been developed and used to study animal social structure in relation to various topics, and the area of network modelling is seeing much current development in the animal behaviour field. The topics include spreading processes [60, 145], cultural diversity [146], cooperation [147], inheritance of network links [148], social stability [82], foraging behaviour [149, 150], and data sampling [93, 151].

**Network methodologies.** Research in animal social networks is accompanied by development of specialised technology and methods for the collection and analysis of data from these systems, which draw upon and are related to general networks & complex systems approaches to different extents. General topics where specialised methods have been developed for animal social networks include automated collection of network data [98, 99], constrained permutation models for statistical testing [110, 111, 152-154], social complexity measures [128], and implications of missing data for the reliability of empirical network structures [115, 155, 156, 157].

## 7 Where to find animal social network data

Data on animal social networks are to an increasing extent being made publicly available in online repositories, including Dryad Digital Repository (datadryad.org), Network Repository (networkrepository.com/asn), and Animal Social Network Repository (bansallab.github.io/asnr [158]), allowing for easy access for complex systems researchers who would like to explore and use such data. Although these data are freely available, we would suggest that the researchers who have provided the data are contacted before the data are used in scientific projects. This is not only as a courtesy to the researcher, but also to make sure that the data are useful for the intended purpose. Factors that may be relevant to consider in this regard include for example the methods used for data collection, the time frame over which the data were collected, the type of behaviour used to quantify the network edges, and potential factors that need to be controlled for in the network construction and analysis (see Section 4).

# 8 Conclusion

Animal social network research is currently at a point where it has become a common part of the study of animal behaviour, and where much knowledge has been obtained about animal social structures. At the same time, this research area is seeing many exciting developments, both in terms of research questions, methods and theory. We see exciting prospects for further integration of complex systems thinking and expertise in this development, and for further integration of animal social networks research into the complex systems & networks field. We believe that the best understanding of animal social networks, and the best use of them for understanding complex systems, is gained by combining intricate knowledge about the specific study systems with innovative and rigorous theory, modelling and analysis. We hope with this overview to have provided a springboard for future cross-disciplinary collaborations around animal social networks.

**Acknowledgements**

This work was funded by the Carlsberg Foundation [a Postdoctoral Internationalisation Fellowship to J.B.B.]; the Natural Environment Research Council [NE/S010327/1 to D.P.C. and S.E.]; and the Leverhulme Trust [an Early Career Research Fellowship to S.E.]. We are grateful to Dan Mønster and three reviewers for helpful comments. We thank Lysanne Snijders and Emily Price for kindly providing photographs.